\documentclass[twocolumn,superscriptaddress,nobalancelastpage,10pt,pra,a4paper]{revtex4}
\usepackage{amsmath}
\usepackage{amsthm}
\usepackage{amssymb}
\usepackage{amsfonts}
\usepackage{graphicx}
\usepackage{wasysym}
\usepackage{mathrsfs}
\usepackage{yfonts}
\usepackage{bbold}
\usepackage{verbatim}
\usepackage{subfigure}

\newcommand{\ket}[1]{\left| #1 \right\rangle}
\newcommand{\bra}[1]{\left\langle #1 \right|}

\newcommand{\opa}{\widehat{a}}

\newcommand{\opEp}{\widehat{\mathbf{E}}^{(+)}}
\newcommand{\opEn}{\widehat{\mathbf{E}}^{(-)}}
\newcommand{\opEps}{\widehat{E}^{(+)}}
\newcommand{\opEns}{\widehat{E}^{(-)}}

\newcommand{\x}{\mathbf{r},t}

\newcommand{\sinc}[1]{\text{sinc}[#1]}

\renewcommand{\r}{\mathbf{r}}
\renewcommand{\k}{\mathbf{k}}

\begin{document}

\title{Theory of Quantum Imaging with Undetected Photons}

\author{Mayukh Lahiri}
\email{mayukh.lahiri@univie.ac.at} \affiliation{Vienna Center for
Quantum Science and Technology (VCQ), Faculty of Physics,
Boltzmanngasse 5, University of Vienna, Vienna A-1090, Austria.}

\author{Radek Lapkiewicz} \affiliation{Vienna Center for
Quantum Science and Technology (VCQ), Faculty of Physics,
Boltzmanngasse 5, University of Vienna, Vienna A-1090, Austria.}

\author{Gabriela Barreto Lemos} \affiliation{Vienna Center for
Quantum Science and Technology (VCQ), Faculty of Physics,
Boltzmanngasse 5, University of Vienna, Vienna A-1090, Austria.}

\author{Anton Zeilinger}\affiliation{Vienna Center for Quantum Science and
Technology (VCQ), Faculty of Physics, Boltzmanngasse 5, University
of Vienna, Vienna A-1090, Austria.}\affiliation{Institute for
Quantum Optics and Quantum Information, Austrian Academy of
Sciences, Boltzmanngasse 3, Vienna A-1090, Austria.}

\begin{abstract}
\noindent \emph{Abstract}: A novel quantum imaging technique has
recently been demonstrated in an experiment, where the photon used
for illuminating an object is not detected; the image is obtained by
interfering two beams, none of which ever interacts with the object.
Here we present a detailed theoretical analysis of the experiment.
We show that the object information is present only in the
interference term and not in the individual intensities of the
interfering beams. We also theoretically establish that the
magnification of the imaging system depends on two wavelengths: the
average wavelength of the photon that illuminates the object and the
average wavelength of the photon that is detected. Our analysis
affirms that the imaging process is based on the principle that
quantum interference occurs when interferometric path information is
unavailable.
\end{abstract}


\maketitle

\section{Introduction}\label{sec:introduction} \noindent
According to Bohr, comprehending the nature of a quantum system
requires ``a combined use of the contrasting pictures'' of a
classical particle and a classical wave \cite{Bohr-Einstein}. Bohr's
complementarity principle \cite{Bohr-comp} implies that the complete
particle behavior and the complete wave behavior of a quantum system
or entity are mutually exclusive. In other words, if a quantum
entity behaves completely like a particle (wave) under certain
experimental conditions, it does not display its wave (particle)
behavior under the same conditions. To avoid confusion, we do not
refer to a quantum entity as either ``particle'' or ``wave'';
instead, we use the term ``quanton'' (see, for example,
\cite{quanton-ref}).
\par
The wave-particle duality can be illustrated by a lowest-order
\cite{Note-order-interf} interference experiment (e.g., Young's
double-slit experiment, Mach-Zehnder interferometer, etc.), in which
a single quanton (e.g., photon, electron, etc.) is sent into a
two-way interferometer (see, for example, \cite{Feynman-lec-3}). If
the quanton behaves completely like a particle, no interference can
be observed at the output of the interferometer. It turns out that
in this case it is possible to determine with complete certainty via
which path the quanton has traversed. On the other hand, when there
is absolutely no information on the path traveled by the quanton,
perfect interference occurs\textemdash a behavior that characterizes
waves. The relationship between interference and path information
(wave-particle duality) has drawn the attention of several
researchers (see, for example,
\cite{GY-neutron-interf,M-ind,JSV,Eng-WPI,ANVKZZ-C60}).
\par
The imaging process \cite{LBCRLZ-mandel-im} of our interest is
related to the wave-particle duality of photons. Let us consider two
spatially separated identical light sources, 1 and 2, each of which
has the ability of producing two photons at a time. These two
photons are, in general, not identical with each other and we label
them by $a$ and $b$. Suppose now that we select the $a$-photons from
the both sources and send them into a two-arm interferometer under
the following conditions: 1) photons from a particular source can
travel through only one of the arms; 2) the sources emit at the same
rate but in such a way that there is never more than one photon
present in the interferometer at a time. In this case, although the
$a$-photons are identical with each other, one can partially or
fully extract the interferometric path information by interacting
with a $b$-photon that is not sent into the interferometer. In such
a situation, it is, therefore, possible to control the interference
of a photon sent into the interferometer by using a photon that is
not sent into the interferometer. This phenomenon has been
experimentally demonstrated and discussed in Refs.
\cite{ZWM-ind-coh-PRL,WZM-ind-coh-PRA} and is often referred to as
``induced coherence without induced emission''.
\par
The essence of our imaging technique \cite{LBCRLZ-mandel-im} lies in
the fact that the effect of interaction with $b$-photons is observed
in the first-order interference fringe pattern produced by the
$a$-photons. As for sources, we use two identical nonlinear crystals
which generate photons by spontaneous parametric down-conversion. In
Section \ref{sec:recol-bas-res}, we briefly recapitulate some basic
results relating to the theory of spontaneous parametric
down-conversion. In Section \ref{sec:imaging}, we then present a
detailed analysis of the imaging method. Finally, we summarize our
results in Section \ref{sec:conclusions}.

\section{Elements of the Theory of Spontaneous Parametric Down-conversion}
\label{sec:recol-bas-res} \noindent We mostly follow the theory of
the process of spontaneous parametric down-conversion developed by
Hong and Mandel \cite{HM-pdc-PRA}. In this process a nonlinear
crystal converts a photon (pump) into two photons (signal and idler)
each of which has energy lower than that of the pump-photon. The
combined energy of the signal and the idler photons is equal to the
energy of the pump-photon. When the pump beam is highly coherent and
the down-conversion does not bring any observable change in the pump
intensity, one can represent the pump by a classical electric field
$\mathbf{E}_P(\x)$. In this case, the interaction Hamiltonian
associated with the process of parametric down-conversion can be
expressed in the interaction picture as (cf.
\cite{WZM-ind-coh-PRA,HM-pdc-PRA})
\begin{align}\label{Ham-pdc}
&\widehat{H}_{in}(t) \nonumber \\ & =\int_D d^3r~
\mbox{\Large$\widetilde{\chi}$}_{lmq}E_{Pl}(\x)
\opEns_{Sm}(\x)\opEns_{Iq}(\x)+\text{H.c.}~,
\end{align}
where $\mbox{\Large$\widetilde{\pmb{\chi}}$}$ represents the
nonlinear electric susceptibility tensor of the crystal,
$\opEn_{S}(\x)$ and $\opEn_{I}(\x)$ are the negative frequency parts
of the quantized electric fields associated with the signal and
idler, respectively, $D$ is the volume of the crystal, H.c. implies
Hermitian conjugation, and there is summation over the repeated
indices $l$, $m$, $q$ which label three mutually orthogonal
directions in space.
\par
The pump, the signal and the idler fields may oscillate at different
optical frequencies. In general, the susceptibility of the crystal
depends on these frequencies. The Hamiltonian in Eq. (\ref{Ham-pdc})
is therefore often expressed by decomposing the optical fields into
several modes (see, for example, \cite{HM-pdc-PRA}). The positive
frequency part of a quantized electric field inside the crystal can
be represented by the expression \cite{Note-em-quant-NLmed}
\begin{align}\label{quant-E-crys-decom}
\opEp{(\x)}&=\sum_{\k,\sigma} \alpha(\k,\sigma) \exp\left[i(\k \cdot
\r -\omega t)\right] \mathbf{e}_{\k,\sigma}~ \opa(\k,\sigma),
\end{align}
where $\sigma=1,2$, labels two directions of polarization, $\omega$
is the frequency, $\k$ is the wave vector, $\mathbf{e}_{\k,\sigma}$
represents two generally complex, mutually orthogonal unit vectors
such that $\mathbf{e}_{\k,\sigma} \cdot \k=0$,
$\alpha(\k,\sigma)=i\sqrt{\hbar \omega/(2\epsilon_0 n^2(\k,\sigma)
L^3)}$, $\epsilon_0$ is the electric permittivity of free space,
$n(\k,\sigma)$ is the refractive index of the anisotropic, nonlinear
crystal, $L^3$ is the quantization volume, and $\opa(\k,\sigma)$ is
the photon annihilation operator for the mode labeled by
$(\k,\sigma)$. Let us also decompose the pump field inside the
crystal into plane wave modes and express it in the form
\begin{align}\label{class-E-decom}
\mathbf{E}_P{(\x)}=\sum_{\k_P,\sigma_P} V_P(\k_P,\sigma_P)
\exp\left[i(\k_P \cdot \r -\omega_P t)\right]
\mathbf{e}_{\k_P,\sigma_P}.
\end{align}
The Hamiltonian, given by Eq. (\ref{Ham-pdc}), now takes the form
(cf. \cite{GHOM-pdc-intf-PRA})
\begin{align}\label{Ham-pdc-sf}
&\widehat{H}_{in}(t) \nonumber \\&=\int_D d^3r \sum_{\k_P,\sigma_P}
\sum_{\k_S,\sigma_S} \sum_{\k_I,\sigma_I}
\Big\{\mbox{\Large$\chi$}_{lmq}(\omega_P,\omega_S,\omega_I)
V_{P}(\k_P,\sigma_P) \nonumber \\& \qquad
\left(\mathbf{e}_{\k_P,\sigma_P}\right)_l
\left(\mathbf{e}^{\ast}_{\k_S,\sigma_S}\right)_m
\left(\mathbf{e}^{\ast}_{\k_I,\sigma_I}\right)_q
\alpha^{\ast}(\k_S,\sigma_S)\alpha^{\ast}(\k_I,\sigma_I) \nonumber
\\ & \qquad \exp\left[i(\omega_S+\omega_I-\omega_P)t\right] \exp
\left[i(\k_P-\k_S-\k_I)\cdot \r \right] \nonumber \\ & \qquad
\opa^{\dag}_S(\k_S,\sigma_S)~\opa^{\dag}_I(\k_I,\sigma_I)\Big\}\quad
+\quad \text{H.c.}~,
\end{align}
where the subscripts $P$, $S$ and $I$ refer to pump, signal and
idler, respectively. The quantum state of light generated by
down-conversion at the crystal is given by the well known formula
\begin{equation}\label{state-pdc-int}
\ket{\psi(t')}=\exp\left[\frac{1}{i\hbar}\int_0^{t'} dt
~\widehat{H}_{in}(t)\right]\ket{\text{vac}},
\end{equation}
where $\ket{\text{vac}}$ is the vacuum state and $t'$ is the
interaction time. By expanding the exponential, Eq.
(\ref{state-pdc-int}) can be expressed in the form
\begin{equation}\label{state-pdc-int-series}
\ket{\psi(t')}=\ket{\text{vac}}+\left[\frac{1}{i\hbar}\int_0^{t'} dt
~\widehat{H}_{in}(t)\right]\ket{\text{vac}}+\dots.
\end{equation}

\section{Imaging} \label{sec:imaging} 
\begin{figure}[htbp]  \centering
  \includegraphics[scale=0.5]{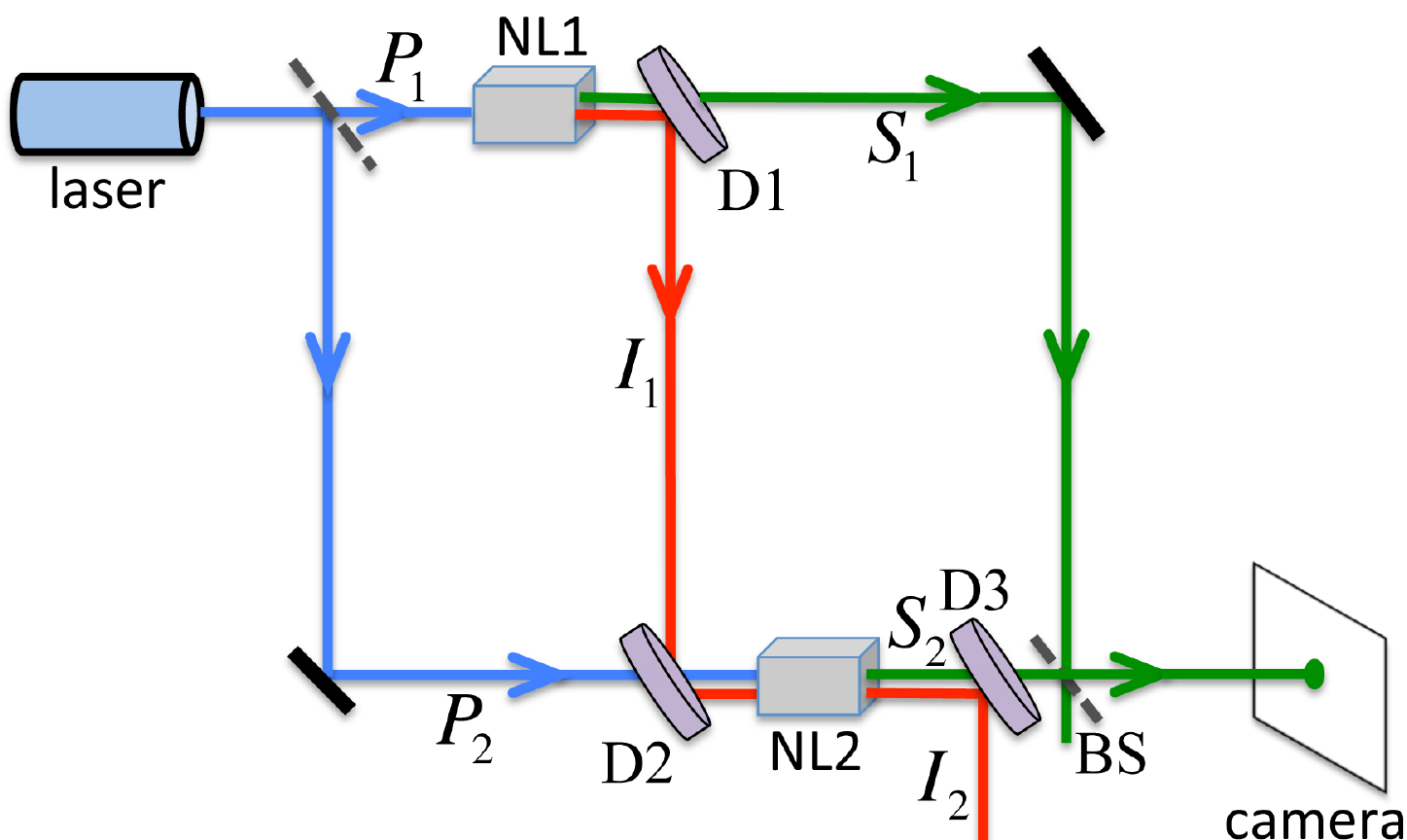}
  \qquad
\caption{Illustrating the principle of the experiment. A laser beam
(blue) is split into two beams $P_1$ and $P_2$ which pump the
nonlinear crystals NL1 and NL2. The crystals produce signal (green)
and idler (red) beams. The idler beam $I_1$ is aligned with the
idler beam $I_2$. The signal beams $S_1$ and $S_2$ are superposed by
the beam splitter BS and one of the outputs of BS is detected by a
camera. D1, D2, D3 are dichroic mirrors.} \label{fig:set-up-schm-ed}
\end{figure}

\noindent
Let us consider a situation in which two identical nonlinear
crystals NL1 and NL2 are pumped by optical beams $P_1$ and $P_2$,
respectively, generated by the same laser source (Fig.
\ref{fig:set-up-schm-ed}).
The idler beam, $I_1$, generated by NL1 is transmitted through NL2
and is aligned with the idler beam, $I_2$, generated by the latter.
The signal beams $S_1$  and $S_2$ from the crystals NL1 and NL2,
respectively, are superposed by a beam-splitter, BS. One of the
outputs of the beam-splitter is detected by an EMCCD camera.
\par
Suppose that the pump fields at the two crystals are given by the
complex electric field vectors $\mathbf{E}_{P_1}(\r_{1},t_{2})$ and
$\mathbf{E}_{P_2}(\r_{2},t_{2})$, expanded in the form given by Eq.
(\ref{class-E-decom}). From Eqs. (\ref{quant-E-crys-decom}),
(\ref{class-E-decom}), (\ref{Ham-pdc-sf}) and
(\ref{state-pdc-int-series}) it follows that the quantum state of
light generated by each individual crystal is given by the formula
(cf. \cite{WMPS})
\begin{align}\label{state-pdc-gen}
&\ket{\psi_j(t')} \nonumber \\ & =\ket{\text{vac}}
+\frac{t'D}{i\hbar} \sum_{\k_{P_j},\sigma_{P_j}}
\sum_{\k_{S_j},\sigma_{S_j}} \sum_{\k_{I_j},\sigma_{I_j}} \Big[
\mbox{\Large$\chi$}_{lmq}(\omega_{P_j};\omega_{S_j},\omega_{I_j})
\nonumber \\ & \quad
V_{P_j}(\k_{P_j},\sigma_{P_j})~\alpha^{\ast}(\k_{S_j},\sigma_{S_j})
\alpha^{\ast}(\k_{I_j},\sigma_{I_j})
\left(\mathbf{e}_{\k_{P_j},\sigma_{P_j}}\right)_l \nonumber \\ &
\quad \left(\mathbf{e}^{\ast}_{\k_{S_j},\sigma_{S_j}}\right)_m
\left(\mathbf{e}^{\ast}_{\k_{I_j},\sigma_{I_j}}\right)_q
\exp\left[i(\omega_{S_j}+\omega_{I_j}-\omega_{P_j})t'/2\right]
\nonumber
\\ & \quad
\sinc{(\omega_{S_j}+\omega_{I_j}-\omega_{P_j})t'/2} \nonumber
\\ & \quad \exp\left[
i(\k_{P_j}-\k_{S_j}-\k_{I_j})\cdot \r_{0_j} \right] \nonumber \\ &
\quad \Big\{\prod_{n=1}^3 \sinc{(\k_{P_j}-\k_{S_j}-\k_{I_j})_nl_n/2}
\Big\} \nonumber \\
& \quad
\ket{\k_{S_j},\sigma_{S_j}}_{S_j}~\ket{\k_{I_j},\sigma_{I_j}}_{I_j}\Big]
+\dots,
\end{align}
where $j=1,2$ labels the two crystals,
$\ket{\k_{S_j},\sigma_{S_j}}_{S_j}=
\opa^{\dag}_{S_j}(\k_{S_j},\sigma_{S_j}) \ket{\text{vac}}_{S_j}$,
$\ket{\k_{I_j},\sigma_{I_j}}_{I_j}=
\opa^{\dag}_{I_j}(\k_{I_j},\sigma_{I_j}) \ket{\text{vac}}_{I_j}$,
the volume integration has been carried out assuming the crystal to
be a rectangular parallelepiped \cite{Note-cryst-vol} of sides
$l_1$, $l_2$, $l_3$ with its center located at the point $\r_{0_j}$,
and $\sinc{x}=\sin x/x$; the sinc terms lead to the two well known
phase matching conditions associated with the process of spontaneous
parametric down-conversion.

\subsection{Alignment of Idler Beams}\label{sec:align-idler}
\noindent If the beam $I_1$ is perfectly aligned with the beam
$I_2$, for each mode present in the quantized field $\opEp_{I_1}$ of
$I_1$ there exists an equally populated mode in the quantized field
$\opEp_{I_2}$ of $I_2$. The perfect alignment of the idler beams
can, therefore, be analytically expressed by the following formula:
\begin{align}\label{a-idl-op-rel-no-ob-gen}
\opa_{I_2}(\k_I,\sigma_I)=
\opa_{I_1}(\widetilde{\k}_I,\widetilde{\sigma}_I)
~\exp[i\phi_I(\widetilde{\k}_I,\widetilde{\sigma}_I)],
\end{align}
where the mode $(\widetilde{\k}_I,\widetilde{\sigma}_I)$ is
generated at NL1 and is aligned with the mode $(\k_I,\sigma_I)$ that
is generated at NL2; $\phi_I(\widetilde{\k}_I,\widetilde{\sigma}_I)$
is a phase factor which can be interpreted as the phase gained by
the mode $(\widetilde{\k}_I,\widetilde{\sigma}_I)$ due to
propagation from NL1 to NL2.
\par
The Hamiltonian given by Eq. (\ref{Ham-pdc-sf}) and, consequently,
the state in Eq. (\ref{state-pdc-gen}) are expressed in quite
general forms. We now simplify them using certain assumptions which
are appropriate to our experiment. Let us assume that the signal and
the idler can be treated as beams with uniform linear polarization
both inside and outside of the crystals. In such a situation, we can
drop the summation over the polarization indices $\sigma_{P_j}$,
$\sigma_{S_j}$, $\sigma_{I_j}$ and can also write
\begin{align}\label{f-approx}
\alpha(\k_{S_j},\sigma_{S_j}) \approxeq \alpha_S(\omega_{S_j}),
\qquad \alpha(\k_{I_j},\sigma_{I_j}) \approxeq
\alpha_I(\omega_{I_j}),
\end{align}
One can now replace the annihilation operator $\opa(\k,\sigma)$ with
$\opa(\k)$, the number state $\ket{\k,\sigma}$ with $\ket{\k}$ and
the susceptibility tensor $\mbox{\Large$\chi$}_{lmq}$ with a scalar
quantity $\mbox{\Large$\chi$}$. Furthermore, the alignment condition
given by Eq. (\ref{a-idl-op-rel-no-ob-gen}) reduces to the form
\begin{align}\label{a-idl-op-rel-no-ob-sp}
\opa_{I_2}(\k_I)=\opa_{I_1}(\widetilde{\k}_I)~\exp[i\phi_I(\widetilde{\k}_I)].
\end{align}
It is to be noted that the relationship between $\widetilde{\k}_I$
and $\k_I$ depends on the optical system used for aligning the idler
beams.
\par
\begin{figure}[htbp]  \centering
  \includegraphics[scale=0.45]{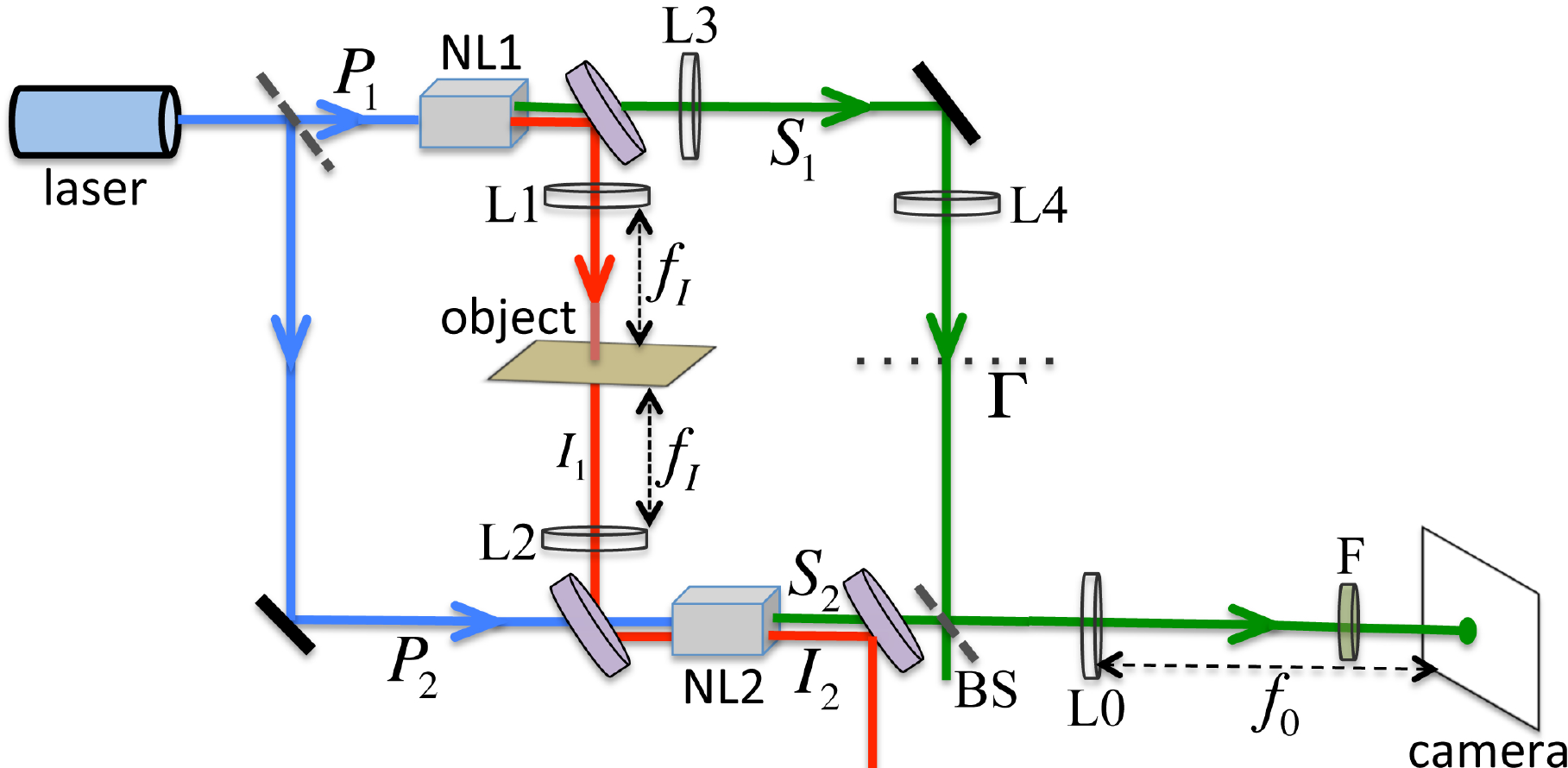}
  \qquad
\caption{Schematics of the imaging experiment. Positive lenses L1
and L2 (both of focal length $f_I$), placed in the path of $I_1$,
form a 4-f system that images NL1 on NL2. An identical 4-f system,
consisting of positive lenses L3 and L4, is placed in the path of
$S_1$. A thin object is placed on a plane which is the back focal
plane of L1 and the front focal plane of L2. One of the outputs of
BS is focused by a positive lens L0 of focal length $f_0$ into a
camera. NL2 and the back focal plane, $\Gamma$, of L4 are located at
the front focal plane of L0. The lenses are assumed to be thin and
ideal. The superposed signal beam is passed through a narrow-band
filter, F, before entering the camera.} \label{fig:set-up-img-ed}
\end{figure}
In the experiment the two idler beams are aligned by the use of a
4-f lens system that images a central plane of NL1 onto a central
plane of NL2 (Fig. \ref{fig:set-up-img-ed}). We assume the idler
beam axis to be along the optical axis of the lens system. A thin
object which is intended to be imaged is placed at the back focal
plane of the first positive lens, L1, of the 4-f system; this plane
is also the front focal plane of the second positive lens, L2, of
the same 4-f system. Clearly, the object is illuminated only by the
idler beam that is generated by the first crystal.
\par
\begin{figure} \centering
 \subfigure[] {
    \label{fig:4-f-obj-ed-a}
    \includegraphics[scale=0.25]{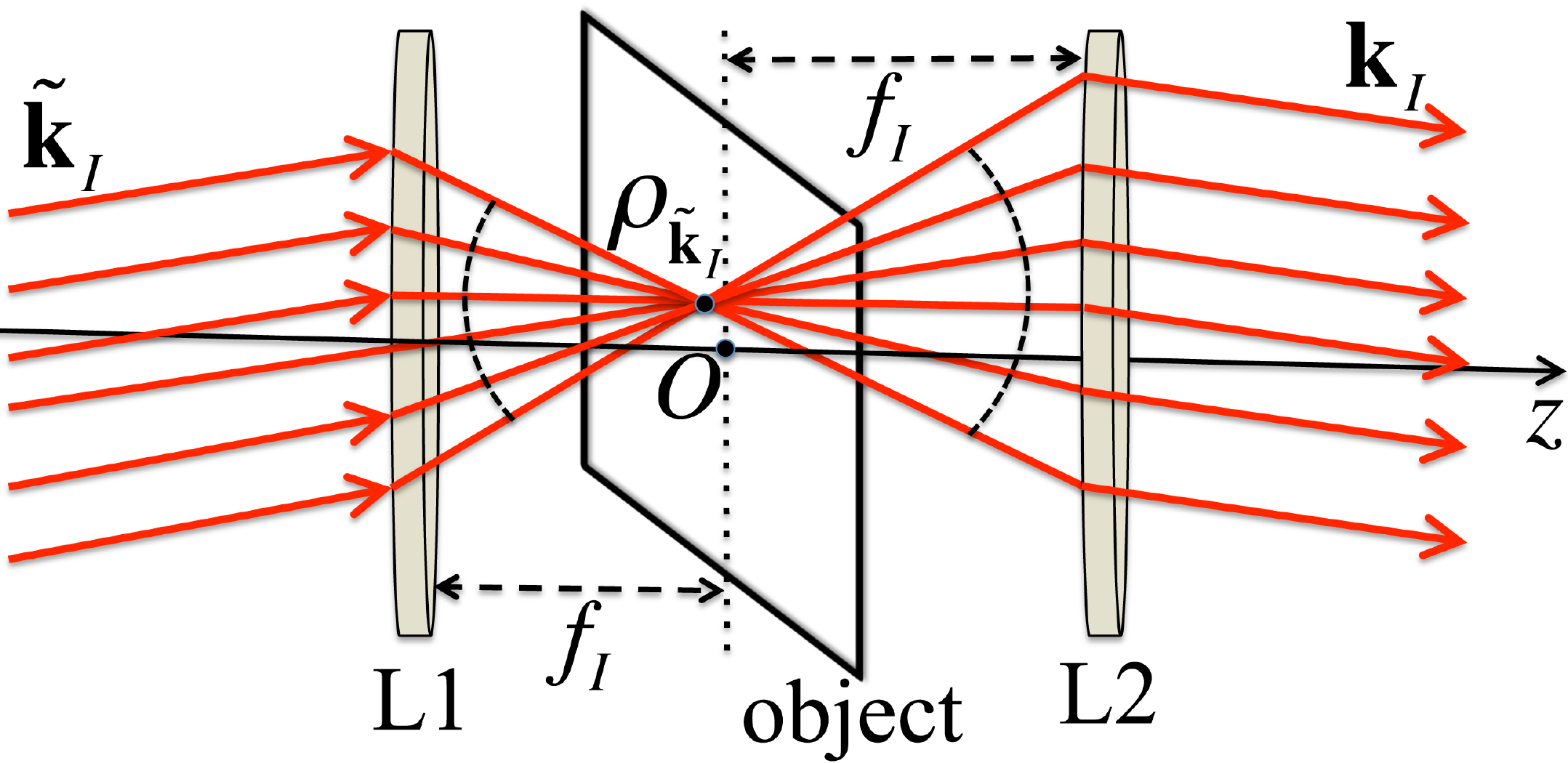}
} \hskip 0.3cm
   \subfigure[] {
    \label{fig:4-f-obj-ed-b}
    \includegraphics[scale=0.33]{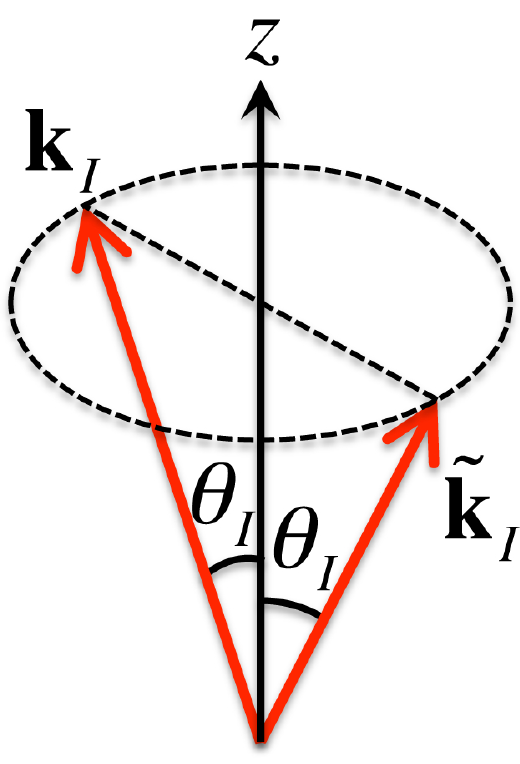}
} \caption{Illustrating notations: (a)
$\pmb{\rho}_{\widetilde{\k}_I}$ is a two-dimensional position vector
lying on the object plane. The origin is chosen at the point $O$
where the optical axis $z$ meets the object plane. We neglect the
limits due to diffraction. (b) The wave vectors $\widetilde{\k}_I$
and $\k_I$ are mirror images of each other with respect to the
optical axis $z$. $\theta_I$ is the absolute value of the angle they
make with the optical axis.} \label{fig:4-f-obj-ed}
\end{figure}

If a plane wave characterized by the wave vector $\widetilde{\k}_I$
is incident on L1, it gets converted into a spherical wave that
converges to a point $\pmb{\rho}_{\widetilde{\k}_I}$, say, on the
back focal plane of L1 [Fig. \ref{fig:4-f-obj-ed-a}]. It then
reemerges from this point as a diverging spherical wave. The
amplitude of the diverging spherical wave can be determined from the
amplitude of the incident wave and the complex transmission
coefficient $\mathscr{T}(\pmb{\rho}_{\widetilde{\k}_I})$ of the
object at point $\pmb{\rho}_{\widetilde{\k}_I}$. The diverging
spherical wave gets reconverted into a plane wave by the positive
lens L2. This plane wave is characterized by a wave vector $\k_I$
which is different from $\widetilde{\k}_I$, unless
$\widetilde{\k}_I$ is along the optical axis $z$ of the lens system.
If one neglects the limits due to diffraction, one can say that a
plane wave emerging from L2 contains information about one specific
point of the object.
\par
Although the discussion of the previous paragraph is applicable to a
classical field, it provides a guideline for treating the problem
quantum mechanically. Since a quantized field is represented by
decomposing it into several plane wave modes, one can say that one
point on the object can transmit and reflect only one specific mode
of the quantized idler field. Hence a single point on the object
acts as a beam splitter only on one particular idler mode. Using the
quantum mechanical treatment of a beam splitter (\cite{MW}, sec.
12.12), one can now write the alignment condition [Eq.
(\ref{a-idl-op-rel-no-ob-sp})] in the following form:
\begin{align}\label{a-idl-op-rel-sp-1}
\opa_{I_2}(\k_I)=&\left[\mathscr{T}(\pmb{\rho}_{\widetilde{\k}_I})
\opa_{I_1}(\widetilde{\k}_I)
+\mathscr{R}'(\pmb{\rho}_{\widetilde{\k}_I})\opa_{0}(\widetilde{\k}_I)
\right] \nonumber
\\ & ~\times \exp[i\phi_I(\widetilde{\k}_I)],
\end{align}
where $\mathscr{T}(\pmb{\rho}_{\widetilde{\k}_I})$ is the
transmission coefficient of the object at the point
$\pmb{\rho}_{\k_I}$, $\mathscr{R}'(\pmb{\rho}_{\widetilde{\k}_I})$
is the reflection coefficient at the same point when illuminated
from the opposite direction, $\opa_{0}$ represents the vacuum field
at the unused port of the beam splitter (a point on the object),
$\phi_I(\widetilde{\k}_I)$ is the phase term mentioned earlier and
$|\mathscr{T}|^2+|\mathscr{R}'|^2=1$. It is evident that in absence
of the object, i.e., when $\mathscr{T}=1$ and $\mathscr{R}'=0$, Eq.
(\ref{a-idl-op-rel-sp-1}) reduces to Eq.
(\ref{a-idl-op-rel-no-ob-sp}).
\par
If one neglects the limits due to diffraction, there is an
one-to-one correspondence between $\k_I$ and $\widetilde{\k}_I$,
i.e., for every choice of $\k_I$ there is one and only one
$\widetilde{\k}_I$. When the focal lengths of L1 and L2 are equal,
$\k_I$ and $\widetilde{\k}_I$ are mirror images of each other with
respect to the optical axis $z$ [see Fig. \ref{fig:4-f-obj-ed-b}].

\subsection{Detection System}\label{sec:det-sys} \noindent
Let us now consider the detection system used in the experimental
setup. The signal beams generated by the two crystals are superposed
by a $50:50$ beam splitter (BS) and one of the outputs of the beam
splitter is focused on an EMCCD camera by a positive lens L0 (Fig.
\ref{fig:set-up-img-ed}). A filter, F, is placed in front of the
camera so that the light entering the camera has a narrow frequency
band of mean frequency $\bar{\omega}_S$.
\begin{figure}[htbp]  \centering
\includegraphics[scale=0.3]{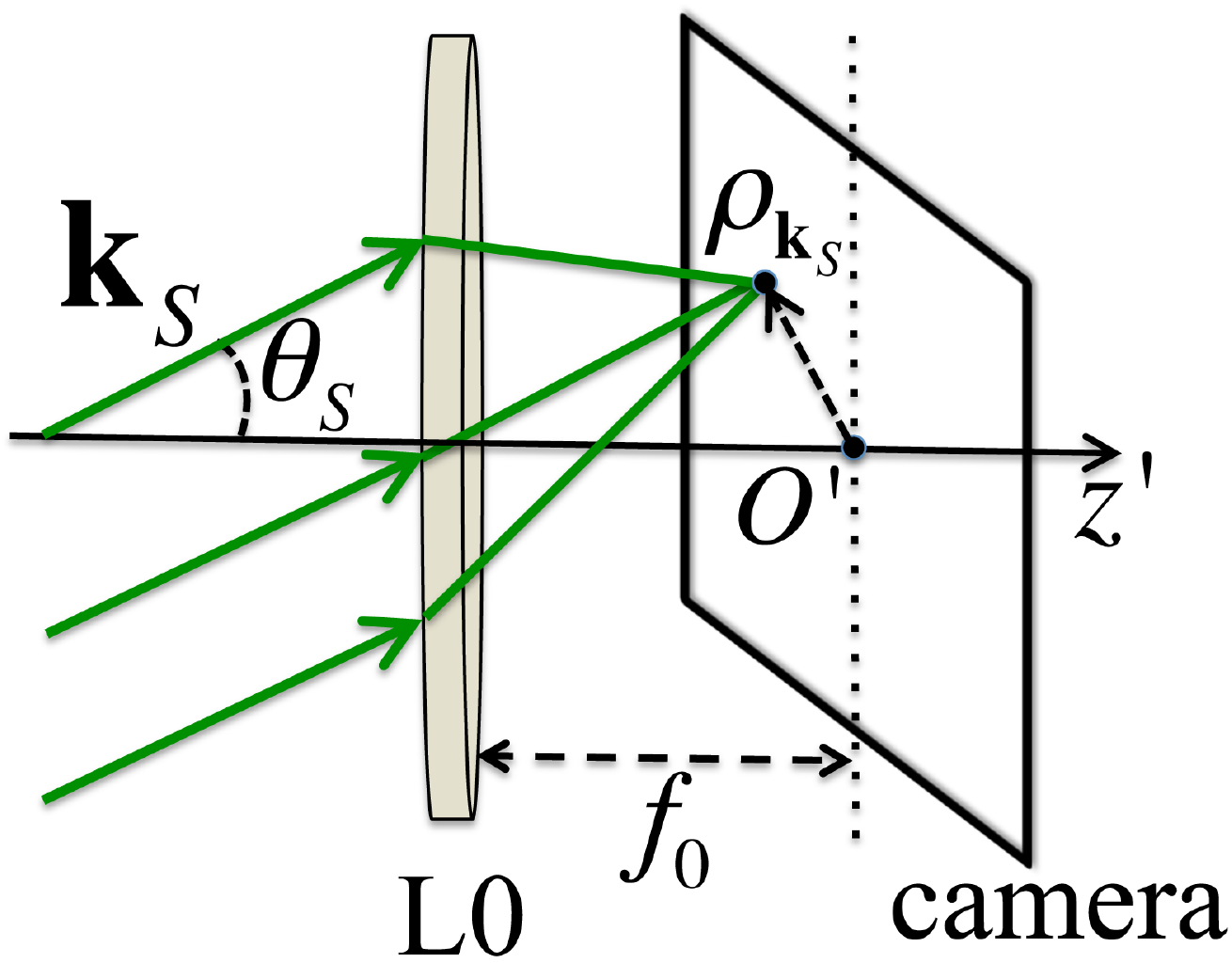}
\qquad \caption{Illustrating notations relating to the detection
system: $\pmb{\rho}_{\k_S}$ is a two-dimensional position vector
lying on the image plane (camera screen). The origin is chosen at
the point $O'$ where the optical axis (also the beam axis) $z'$
meets the image plane. The wave vector $\k_S$ makes an angle
$\theta_S$ with the optical axis.} \label{fig:det-sys-ed}
\end{figure}
\par
In absence of limits due to diffraction, L0 would focus a plane wave
with wave vector $\k_S$ at a point $\pmb{\rho}_{\k_S}$, say, on the
camera screen (Fig. \ref{fig:det-sys-ed}). Hence the positive
frequency part of the quantized field at the camera can be expressed
as \cite{Note-cryst-refrac}
\begin{align}\label{E-quant-at-cam-nb-im}
&\opEps_{S}(\pmb{\rho}_{\k_S},t) \nonumber \\ & \propto
\exp\left[-i\bar{\omega}_S\left(t-\frac{L_1(\k_S)}c\right)
\right]~\opa_{S_1}(\k_{S}) \nonumber \\ & \quad +i
\exp\left[i\k_S\cdot
\r_0-i\bar{\omega}_S\left(t-\frac{L_2(\k_S)}c\right)
\right]~\opa_{S_2}(\k_{S}),
\end{align}
where $L_j(\k_S)$ is the optical path traveled by the mode labeled
by $\k_S$ from the nonlinear crystal $j$ to the camera (propagation
inside the crystals has been neglected), $|\k_S|=\bar{\omega}_S/c$
and we have chosen $\r_{0_1}=0$, $\r_{0_2}-\r_{0_1}=\r_0$.

\subsection{Formation of an Image}\label{sec:theory-imaging} \noindent
For the sake of simplicity, we assume that the pump beams are well
collimated, uniformly polarized and narrow-band with mean frequency
$\bar{\omega}_P$. In this case, the pump field at the crystal $j$
can be represented by $V_{P_j}\exp[i(\k_{P} \cdot
\r-\bar{\omega}_Pt)]$. We choose the direction of $\k_P$ to be along
the direction of the optical axis. Using Eqs. (\ref{state-pdc-gen}),
(\ref{f-approx}) and (\ref{a-idl-op-rel-sp-1}) one can show (see
Appendix 1) that the quantum state of the field in this system can
be approximated by
\begin{align}\label{state-final-approx}
&\ket{\Psi} \approx \nonumber \\
& \ket{\text{vac}}+\frac{t'D}{i\hbar} \sum_{\k_{S_1}}
\sum_{\k_{I_1}} \Big[g(\omega_{S_1},\omega_{I_1}) V_{P_1}
\nonumber \\
& \big\{\prod_{n=1}^3\sinc{(\k_{P}-\k_{S_1}-\k_{I_1})_nl_n/2}
\big\} \nonumber \\
& \sinc{(\omega_{S_1}+\omega_{I_1}-\bar{\omega}_P)t'/2}
\ket{\k_{S_1}}_{S_1}
\ket{\k_{I_1}}_{I_1} \ket{~0~}_{S_2}\ket{0}_0 \Big] \nonumber \\
& +\frac{t'D}{i\hbar} \sum_{\k_{S_2}} \sum_{\k_{I_2}} \Big[
g(\omega_{S_2},\omega_{I_2}) V_{P_2} \exp\left[
i(\k_{P}-\k_{S_2}-\k_{I_2})\cdot \r_0 \right] \nonumber \\ &
\exp[-i\phi_I(\widetilde{\k}_{I_2})]
\sinc{(\omega_{S_2}+\omega_{I_2}-\bar{\omega}_P)t'/2} \nonumber \\
& \big\{\prod_{n=1}^3\sinc{(\k_{P}-\k_{S_2}-\k_{I_2})_nl_n/2} \big\}
\nonumber \\ & ~\Big(
\mathscr{T}^{\ast}(\pmb{\rho}_{\widetilde{\k}_{I_2}})\ket{~0~}_{S_1}
~\ket{\widetilde{\k}_{I_2}}_{I_1} \ket{\k_{S_2}}_{S_2}\ket{0}_0
\nonumber \\ & ~
+\mathscr{R}^{'\ast}(\pmb{\rho}_{\widetilde{\k}_{I_2}})
\ket{~0~}_{S_1}~\ket{0}_{I_1}
\ket{\k_{S_2}}_{S_2}\ket{\widetilde{\k}_{I_2}}_0 \Big) \Big],
\end{align}
where $g(\omega_{S_j},\omega_{I_j})=\mbox{\Large$\chi$}
(\omega_{S_j},\omega_{I_j})\alpha^{\ast}_S(\omega_{S_j})
\alpha^{\ast}_I(\omega_{I_j})\exp [i(\omega_{S_j}
+\omega_{I_j}-\bar{\omega}_P)t'/2]$, $\widetilde{\k}_{I_2}$ is the
mirror image of $\k_{I_2}$ with respect to the optical axis (beam
axis) of the 4-f system and we have suppressed the normalization
coefficients. In the experiment, we choose crystals whose sides are
approximately $10^{-3}$m of length; since the wave vectors are
characterized by corresponding optical wavelengths, the terms
$\sinc{(\k_{P_j}-\k_{S_j}-\k_{I_j})_nl_n/2}$ in Eq.
(\ref{state-pdc-gen}) contribute only when
$\k_{P_j}-\k_{S_j}-\k_{I_j}\approx 0$. This implies that the spatial
phase matching condition $\k_{P_j} \approx \k_{S_j}+\k_{I_j}$ holds
to a very good accuracy. Furthermore, the term
$\sinc{(\omega_{S_j}+\omega_{I_j}-\bar{\omega}_P)t'/2}$ leads to the
temporal phase matching condition $\bar{\omega}_P\approx
\omega_{S_j}+\omega_{I_j}$.
\par
The photon counting rate \cite{G-1} at a point $\pmb{\rho}_{\k_S}$
in the camera is given by
\begin{equation}\label{R-at-cam-def}
\mathcal{R}(\pmb{\rho}_{\k_S}) \propto
\bra{\Psi}\opEns_{S}(\pmb{\rho}_{\k_S},t)\opEps_{S}(\pmb{\rho}_{\k_S},t)
\ket{\Psi},
\end{equation}
where $\ket{\Psi}$ is given by Eq. (\ref{state-final-approx}) and
the quantized field is given by Eq. (\ref{E-quant-at-cam-nb-im}). It
follows from a long but straightforward calculation that apart from
a proportionality constant $\mathcal{R}(\pmb{\rho}_{\k_S})$ is given
by
\begin{align}\label{R-Ks-at-cam-nbp-im}
\mathcal{R}(\pmb{\rho}_{\k_S})& \approx |V_{P_1}|^2 +|V_{P_2}|^2 \nonumber \\
& +2|V_{P_1}||V_{P_2}||\mathscr{T}(\pmb{\rho}_{\widetilde{\k}_I})| \nonumber \\
& \qquad \cos\Big[\phi_{S_2}(\k_{S})
-\phi_{S_1}(\k_{S})-\phi_I(\widetilde{\k}_{I}) \nonumber \\
& \qquad \qquad
-\text{arg}[\mathscr{T}(\pmb{\rho}_{\widetilde{\k}_I})] +\phi_P
+\k_S\cdot \r_0+C_0 \Big],
\end{align}
where $\widetilde{\k}_{I}$ denotes a wave vector that is the mirror
image of the wave vector $\k_I=\k_P-\k_S$ with respect to the
optical axis [Fig. \ref{fig:4-f-obj-ed-b}],
$|\k_P|=\bar{\omega}_P/c$, $|\k_S|=\bar{\omega}_S/c$,
$|\k_I|=\bar{\omega}_I/c$, $\pmb{\rho}_{\widetilde{\k}_I}$ is the
point on the object that is illuminated by the idler mode
$\widetilde{\k}_{I}$, $\phi_{S_j}(\k_{S})=\bar{\omega}_S
L_j(\k_S)/c$, $\phi_P=\text{arg}[V_{P_2}]-\text{arg}[V_{P_1}]$, and
$C_0$ includes all other phase terms. Equation
(\ref{R-Ks-at-cam-nbp-im}) is the key equation of the theory of
imaging.
\par
Let us first consider the situation in which no object is placed in
the idler's path, i.e., when $|\mathscr{T}|=1$ and
$\text{arg}[\mathscr{T}]=0$. We have already mentioned in Section
\ref{sec:det-sys} that a point on the image plane (camera) has
contribution only from one signal mode $\k_S$. Since the diameter of
the signal beam cross-section in the camera is much smaller in
dimension than the optical paths $L_j(\k_S)$ and the distance
$|\r_0|$ between the two crystals, the terms $\k_S\cdot \r_0$ and
$\phi_{Sj}(\k_{S})$ can be treated as a slowly varying function of
$\k_S$. Similarly, since the diameter of the first idler beam inside
NL2 is much smaller than the distance between the two crystals, one
can also neglect the $\widetilde{\k}_I$ dependence of
$\phi_I(\widetilde{\k}_{I})$. These allow us to write
$\phi_{S_2}(\k_{S}) -\phi_{S_1}(\k_{S})\approx \Delta_{S0}$,
$\phi_I(\widetilde{\k}_{I})\approx \phi_{I0}$ and $\k_S\cdot
\r_0+C_0\approx C'_0$, where $\Delta_{S0}$, $\phi_{I0}$ and $C'_0$
are constants. Equation (\ref{R-Ks-at-cam-nbp-im}) now reduces to
\begin{align}\label{R-img-cond-no-obj}
\mathcal{R}(\pmb{\rho}_{\k_S}) \approx & |V_{P_1}|^2 +|V_{P_2}|^2 \nonumber \\
& +2|V_{P_1}||V_{P_2}|\cos\Big[\Delta_{S0}-\phi_{I0} +\phi_P +C'_0
\Big].
\end{align}
Since the right-hand side of Eq. (\ref{R-img-cond-no-obj}) does not
have any $\k_S$ dependence, it is clear that an almost uniformly
illuminated beam cross-section would be observed in the camera. The
phase term $\phi_P$ can be controlled in the experiment and by doing
so one can modulate the intensity of the beam spot. It is evident
from Eq. (\ref{R-img-cond-no-obj}) that by changing the value of
$\phi_P$, one can achieve conditions both of constructive and of
destructive interference, which are given by the following
equations, respectively:
\begin{subequations}\label{imaging-max-min-cond}
\begin{align}
&\Delta_{S0}-\phi_{I0} +\phi_{PC} +C'_0 =2N\pi,
\label{imaging-max-min-cond:a} \\
&\Delta_{S0}-\phi_{I0} +\phi_{PD} +C'_0 =(2N+1)\pi,
\label{imaging-max-min-cond:b}
\end{align}
\end{subequations}
where $\phi_{PC}$ and $\phi_{PD}$ are values of $\phi_P$ for
constructive and destructive interferences \cite{Note-bs-output},
respectively, and $N$ is an integer. Thus we have established the
relations
\begin{subequations}\label{imaging-cond}
\begin{align}
&\phi_{S_2}(\k_{S})
-\phi_{S_1}(\k_{S})-\phi_I(\widetilde{\k}_{I})+\phi_{PC}
+C'_0\approx 2N\pi,
\label{imaging-cond:a} \\
&\phi_{S_2}(\k_{S})
-\phi_{S_1}(\k_{S})-\phi_I(\widetilde{\k}_{I})+\phi_{PD}
+C'_0\approx (2N+1)\pi. \label{imaging-cond:b}
\end{align}
\end{subequations}
\par
When the object is inserted in the idler's path, it follows from
Eqs. (\ref{R-Ks-at-cam-nbp-im}) and (\ref{imaging-cond}) that the
photon counting rates at a point in the camera under the conditions
of constructive and destructive interference are, respectively,
given by the formulas
\begin{subequations}\label{R-Ks-at-cam-im-pos-neg}
\begin{align}
\mathcal{R}^{(+)}(\pmb{\rho}_{\k_S})& \approx |V_{P_1}|^2
+|V_{P_2}|^2 \nonumber \\ &
+2|V_{P_1}||V_{P_2}||\mathscr{T}(\pmb{\rho}_{\widetilde{\k}_I})|
\cos( \text{arg}[\mathscr{T}(\pmb{\rho}_{\widetilde{\k}_I})]),
\label{R-Ks-at-cam-im-pos-neg:a} \\
\mathcal{R}^{(-)}(\pmb{\rho}_{\k_S})& \approx |V_{P_1}|^2
+|V_{P_2}|^2 \nonumber \\ &
-2|V_{P_1}||V_{P_2}||\mathscr{T}(\pmb{\rho}_{\widetilde{\k}_I})|
\cos( \text{arg}[\mathscr{T}(\pmb{\rho}_{\widetilde{\k}_I})]).
\label{R-Ks-at-cam-im-pos-neg:b}
\end{align}
\end{subequations}
Equations (\ref{R-Ks-at-cam-im-pos-neg}) imply that an image of an
absorptive object ($\text{arg}[\mathscr{T}(\pmb{\rho}_{\k_I})]=0$)
and as well as of a phase object
($|\mathscr{T}(\pmb{\rho}_{\k_I})|=1$) would appear in the camera
for both constructive and destructive interferences (see
\cite{LBCRLZ-mandel-im}, Fig. 3a). It further follows from Eqs.
(\ref{R-Ks-at-cam-im-pos-neg}) that apart from a proportionality
constant
\begin{align}\label{imaging-eq}
\mathcal{R}^{(+)}(\pmb{\rho}_{\k_S})-\mathcal{R}^{(-)}(\pmb{\rho}_{\k_S})
\approx |\mathscr{T}(\pmb{\rho}_{\widetilde{\k}_I})| \cos(
\text{arg}[\mathscr{T}(\pmb{\rho}_{\widetilde{\k}_I})]),
\end{align}
implying the background effect due to presence of the terms
$|V_{P_1}|^2 $ and $|V_{P_2}|^2$ can be eliminated by subtracting
the photon counting rate obtained with destructive interference from
that obtained with constructive interference (see
\cite{LBCRLZ-mandel-im}, Fig. 3d). It also follows from Eqs.
(\ref{R-Ks-at-cam-im-pos-neg}) that
\begin{align}\label{bs-output-sum}
\mathcal{R}^{(+)}(\pmb{\rho}_{\k_S})+\mathcal{R}^{(-)}(\pmb{\rho}_{\k_S})
\approx 2(|V_{P_1}|^2 +|V_{P_2}|^2).
\end{align}
This means that summing up the photon counting rates obtained by
constructive and destructive interferences removes the image (see
\cite{LBCRLZ-mandel-im}, Fig. 3c). Clearly, even if one uses an
absorptive object that completely blocks the beam $I_1$, the
summation of the photon counting rates does not change. The fact
that the information of the object appears only in the interference
term shows that the imaging process is purely quantum mechanical in
nature; this point is discussed later in further details.
\par
It is clear from the preceding discussion that a point
$\pmb{\rho}_{\widetilde{\k}_I}$ in the object plane is imaged at the
point $\pmb{\rho}_{\k_S}$ in the image plane, where
$\pmb{\rho}_{\widetilde{\k}_I}$ is the point at which a classical
plane wave with wave vector $\widetilde{\k}_I$ would be focused by
the lens L1 [see Fig. \ref{fig:4-f-obj-ed-a}] and
$\pmb{\rho}_{\k_S}$ is the point where the plane wave characterized
by $\k_S$ would be focused by L0. Since $\widetilde{\k}_I$ is the
mirror image of the wave vector $\k_I$ with respect to the optical
axis [see Fig. \ref{fig:4-f-obj-ed-b}] and $\k_I$ is related to
$\k_S$ by the phase matching condition $\k_I\approx \k_P-\k_S$, the
image that appears on the camera is not inverted
\cite{Note-img-ninv-lens}.

\subsection{Image Magnification}\label{sec:image-magnification}
\noindent So far we have neglected the effect of refraction at the
crystal surface. However, one needs to consider this effect in order
to obtain a correct value of the magnification. The wave vectors
$\k_S$ and $\widetilde{\k}_I$ used thus far represent plane waves
outside the crystals. The mirror image of $\widetilde{\k}_I$ with
respect to the optical axis $z$ is $\k_I$ [Fig.
\ref{fig:4-f-obj-ed-b}]. Suppose that the plane waves with wave
vectors $\k_S$ and $\k_I$ are represented inside the crystal by
$\k''_S$ and $\k''_I$, respectively. The phase matching condition
can now be expressed as
\begin{align}\label{phase-mt-wvec}
\k''_S+\k''_I\approx \k''_P,
\end{align}
where $\k''_P$ is the wave vector of the pump field inside the
crystal. If $\k''_S$ and $\k''_I$ make angles $\theta''_S$ and
$\theta''_I$, respectively, with $\k''_P$ which is along the optical
axis, it follows from Eq. (\ref{phase-mt-wvec}) that
\begin{align}\label{phase-mt-angle}
\bar{\omega}_S n_S |\sin\theta''_S|\approx \bar{\omega}_I n_I
|\sin\theta''_I|.
\end{align}
\par
Let us choose two points on the object which are represented by
two-dimensional position vectors $\pmb{\rho}_{\widetilde{\k}_I}$ and
$\pmb{\rho}'_{\widetilde{\k}'_I}$. Suppose that their images at the
camera are represented by the two-dimensional position vectors
$\pmb{\rho}_{\k_S}$ and $\pmb{\rho}'_{\k'_S}$, respectively. The
magnification is defined by the well known formula
\begin{align}\label{mag-def}
M=\frac{|\pmb{\rho}_{\k_S}-\pmb{\rho}'_{\k'_S}|}
{|\pmb{\rho}_{\widetilde{\k}_I}-\pmb{\rho}'_{\widetilde{\k}'_I}|},
\end{align}
where its positive sign implies that the image is not inverted. As
already mentioned, the origins in the object and the image planes
are chosen at the points $O$ and $O'$ where the corresponding
optical axes (beam axes) meet the respective planes [see Figs.
\ref{fig:4-f-obj-ed-a} and \ref{fig:det-sys-ed}]. It readily follows
from the theory presented in Section \ref{sec:theory-imaging} that
$O'$ is the image of $O$. By choosing the points
$\pmb{\rho}'_{\widetilde{\k}'_I}$ and $\pmb{\rho}'_{\k'_S}$ at $O$
and $O'$ respectively, we reduce Eq. (\ref{mag-def}) to the
simplified form
\begin{align}\label{mag-def-simpl}
M=\frac{|\pmb{\rho}_{\k_S}|}{|\pmb{\rho}_{\widetilde{\k}_I}|}.
\end{align}
If the signal plane wave $\k_S$, which is focused by the lens L1 at
the point $\pmb{\rho}_{\k_S}$, makes an angle $\theta_S$ with the
optical axis $z'$ (Fig. \ref{fig:det-sys-ed}), one has in the
small-angle approximation
$|\pmb{\rho}_{\k_S}|=|f_0\tan\theta_S|\approx f_0|\theta_S|$.
Similarly, one can show that $|\pmb{\rho}_{\widetilde{\k}_I}|\approx
f_I |\theta_I|$, where $\theta_I$ is the angle made by the wave
vector $\widetilde{\k}_I$ with the optical axis [Fig.
\ref{fig:4-f-obj-ed-a}]. It now follows from Eq.
(\ref{mag-def-simpl}) that
\begin{align}\label{mag-formula}
M=\frac{f_0 |\theta_S|}{f_I |\theta_I|}.
\end{align}
\par
Since $\k_S$ and $\k_I$ are related to $\k''_S$ and $\k''_I$,
respectively, by refraction at the crystal surface, using Snell's
law one obtains
\begin{align}\label{Snell-law-crys}
n_S\sin\theta''_S=n_0\sin\theta_S, \quad
n_I\sin\theta''_I=n_0\sin\theta_I,
\end{align}
where we have used the fact that $\k_I$ and $\widetilde{\k}_I$ make
the same angle with the optical axis [Fig. \ref{fig:4-f-obj-ed-b}]
and the refractive index of air ($n_0$) has practically the same
value for signal and idler. From Eqs. (\ref{phase-mt-angle}) and
(\ref{Snell-law-crys}), it immediately follows that $\bar{\omega}_S
|\sin\theta_S| \approx \bar{\omega}_I |\sin\theta_I|$. In the
small-angle limit, we thus obtain
\begin{align}\label{freq-ang-rel}
\bar{\omega}_S|\theta_S|\approx \bar{\omega}_I|\theta_I|.
\end{align}
From Eqs. (\ref{mag-formula}) and (\ref{freq-ang-rel}) one finds
that
\begin{align}\label{mag-result}
M=\frac{f_0 \bar{\omega}_I}{f_I \bar{\omega}_S} =\frac{f_0
\bar{\lambda}_S}{f_I \bar{\lambda}_I}.
\end{align}
Clearly, magnification of the imaging system depends on the ratio of
the average wavelength of the signal to that of the idler. The
dependence of the magnification on both wavelengths is a remarkable
feature of our imaging process.

\section{Conclusion}\label{sec:conclusions}
\noindent We have theoretically analyzed a recently demonstrated
\cite{LBCRLZ-mandel-im} quantum imaging technique. Although the
experimental setup resembles an ordinary two-arm interferometer, the
principle behind the imaging is purely quantum mechanical. If one
imagines a classical two-arm interferometer in one of whose arms an
absorptive object is placed, both the interference term and the
intensity contribution from the arm containing the object would
depend on the transmissivity of the object. In our experiment, on
the other hand, the intensity contribution from any of the crystals
does not depend on the transmissivity of the object. It is evident
from Eqs. (\ref{R-Ks-at-cam-im-pos-neg}) that the information of the
object is present \emph{only} in the interference term. This also
shows that the interference of signal beams is not due to the effect
of induced emission (see also
\cite{ZWM-ind-coh-PRL,WM-ind-coh-cl-q}). This interference can only
be explained by indistinguishability of the signal photon paths and
hence the imaging process is directly related to wave-particle
duality of photons.
\par
In this context, let us also have a close look at Eq.
(\ref{state-final-approx}), which provides us with an expression for
the quantum state $\ket{\Psi}$ that has been used for explaining the
imaging process. Since the pump source used in the experiment is a
narrow-band laser, this state is obtained (see Appendix 1) from the
tensor product of the individual states, $\ket{\psi_1}$ and
$\ket{\psi_2}$, generated by the crystals under the
alignment-condition imposed by Eq. (\ref{a-idl-op-rel-sp-1}). This
tensor product together with the alignment-condition implies the
effect of induced emission. However, when the higher order terms
present in $\ket{\psi_1}$ and $\ket{\psi_2}$ can be neglected, the
state $\ket{\Psi}$ becomes identical with a state obtained by linear
superposition of $\ket{\psi_1}$ and $\ket{\psi_2}$ under the same
alignment-condition (see Appendix 1). Since such a superposition is
only allowed when there is no effect of induced emission, it is
clear that photons generated by spontaneous parametric down
conversion play the key role in our imaging process
\cite{Note-ind-em}.
\par
Finally, although it is obvious, we would like to point out that the
principle of our imaging also works if an entirely different lens
system (or no lens system) is used in the experiment. A different
lens system would only lead to a different value of the image
magnification. Equation (\ref{mag-result}) shows that this
magnification is equal to the product of two ratios: the ratio of
focal lengths ($f_0/f_I$) and the ratio of wavelengths
($\bar{\lambda}_S/\bar{\lambda}_I$). The most remarkable feature of
this result is the presence of two mean wavelengths in the formula
of magnification. This is a consequence of (a) the fact that the
object is illuminated by photons of one mean wavelength while the
camera detects photons of the other wavelength and (b) the
phase-matching condition [see Eq. (\ref{phase-mt-angle})]. The use
of a different lens system in the setup might lead to a different
expression for the image magnification; however, a dependence on two
average wavelengths would always be present.

\section*{Acknowledgements}
The authors thanks M. Horne for many discussions. This project was
supported by \"OAW, the European Research Council (ERC Advanced
grant no. 227844 ``QIT4QAD'', and SIQS grant no. 600645 EU-FP7-ICT),
and the Austrian Science Fund (FWF) with SFB F40 (FOQUS) and W1210-2
(CoQus).

\section*{Appendix 1} \noindent
In this appendix, we illustrate the procedure used for obtaining Eq.
(\ref{state-final-approx}). We present a single-mode analysis,
because the procedure does not change when all the modes present in
the quantized fields are considered. The single-mode and scalar
version of Eq. (\ref{state-pdc-gen}) can be represented in the form
\begin{align}\label{state-pdc-gen-sm}
\ket{\psi_j}=\ket{\text{vac}} +G_j \ket{1}_{S_j}~\ket{1}_{I_j}
+\dots,
\end{align}
where $j=1,2$, the time dependence is suppressed, the coefficients
are collectively denoted by $G_j$ and the dots represent higher
order terms containing higher powers of $G_j$.
\par
The single-mode version of Eq. (\ref{a-idl-op-rel-sp-1}) is given by
\begin{align}\label{a-idl-op-rel-sp-sm}
\opa_{I_2}=\left[\mathscr{T} \opa_{I_1} +\mathscr{R}'\opa_{0}
\right]~e^{i\phi_I}.
\end{align}
From Eq. (\ref{a-idl-op-rel-sp-sm}), one immediately obtains that
\begin{align}\label{st-idl-1-2-rel-sm}
\ket{1}_{I_2}=\left[\mathscr{T}^{\ast} \ket{1}_{I_1}\ket{0}_{0}
+\mathscr{R}'^{\ast}\ket{0}_{I_1}\ket{1}_{0} \right]~e^{-i\phi_I}.
\end{align}
Now using Eqs. (\ref{state-pdc-gen-sm}) and
(\ref{st-idl-1-2-rel-sm}), one can write
\begin{subequations}\label{state-pdc-gen-1-2-sm}
\begin{align}
&\ket{\psi_1}=\ket{\text{vac}} +G_1 \ket{1}_{S_1}~\ket{1}_{I_1}
\ket{0}_{S_2}~\ket{0}_{0}
+\dots \label{state-pdc-gen-1-2-sm:a} \\
&\ket{\psi_2}=\ket{\text{vac}} +G_2 e^{-i\phi_I}
\big[\mathscr{T}^{\ast}
\ket{0}_{S_1}\ket{1}_{I_1}\ket{1}_{S_2}\ket{0}_{0} \nonumber \\ &
\qquad \qquad \qquad \qquad +\mathscr{R}'^{\ast}
\ket{0}_{S_1}\ket{0}_{I_1}\ket{1}_{S_2}\ket{1}_{0} \big]+\dots.
\label{state-pdc-gen-1-2-sm:b}
\end{align}
\end{subequations}
\par
Since both crystals are pumped by beams generated by a laser source
(not a single photon source), the quantum state of light in the
system is given by $\ket{\Psi}=\ket{\psi_1}\ket{\psi_2}$, i.e., by
\begin{align}\label{state-pdc-gen-both-sm}
\ket{\Psi}&=\ket{\text{vac}} +G_1 \ket{1}_{S_1}~\ket{1}_{I_1}
\ket{0}_{S_2}~\ket{0}_{I_2} \nonumber \\ & +G_2 e^{-i\phi_I}
\big[\mathscr{T}^{\ast}
\ket{0}_{S_1}\ket{1}_{I_1}\ket{1}_{S_2}\ket{0}_{0} \nonumber \\
& +\mathscr{R}'^{\ast}
\ket{0}_{S_1}\ket{0}_{I_1}\ket{1}_{S_2}\ket{1}_{0} \big] \nonumber
\\ & +\text{higher order terms},
\end{align}
where the higher order terms contains higher second or higher powers
of the coefficients $G$. Since the rate of down conversion is very
small, the these higher order terms can be neglected. Hence the
state $\ket{\Psi}$ can be approximated by (neglecting the
normalization coefficients)
\begin{align}\label{state-pdc-gen-both-sm-approx}
\ket{\Psi}& \approx \ket{\text{vac}} +G_1
\ket{1}_{S_1}~\ket{1}_{I_1} \ket{0}_{S_2}~\ket{0}_{I_2} \nonumber \\
& +G_2 e^{-i\phi_I} \big[\mathscr{T}^{\ast}
\ket{0}_{S_1}\ket{1}_{I_1}\ket{1}_{S_2}\ket{0}_{0} \nonumber \\
& \qquad +\mathscr{R}'^{\ast}
\ket{0}_{S_1}\ket{0}_{I_1}\ket{1}_{S_2}\ket{1}_{0} \big].
\end{align}
Equation (\ref{state-final-approx}) is the multi-mode version of Eq.
(\ref{state-pdc-gen-both-sm-approx}).
\par
It is to be noted that the form of $\ket{\Psi}$ given by Eq.
(\ref{state-pdc-gen-both-sm-approx}) can also be obtained by linear
superposition of the states given by Eqs.
(\ref{state-pdc-gen-1-2-sm:a}) and (\ref{state-pdc-gen-1-2-sm:a}),
if one neglects the higher order terms. It is thus clear that if the
experimental conditions are such that the contribution of these
higher order terms is much smaller than that of the first order
terms to a measurement of our interest, the measurement would yield
the same result as in a case when the state $\ket{\Psi}$ must be
obtained by superposing $\ket{\psi_1}$ and $\ket{\psi_2}$.



\begin{thebibliography}{99}
\bibitem {Note-order-interf} The lowest-order interference has been
referred to as both second-order \cite{MW} and first-order
\cite{G-1} in the literature.
\bibitem {Note-em-quant-NLmed} Quantization of electromagnetic
fields inside a nonlinear crystal is a non-trivial problem. However,
when the nonlinearities are not high, the field is often represented
in a form which is analogous to that of the free quantized field
(see, for example, \cite{MW}, Ch. 22).
\bibitem {Note-cryst-vol} We assume that almost the entire volume of each
crystal is illuminated by the pump beam.
\bibitem {Note-cryst-refrac} The photons are detected outside the
crystals, whereas the phase matching conditions represent criteria
inside the crystals. Hence the signal wave vectors used in Eq.
(\ref{E-quant-at-cam-nb-im}) are related to the ones used in Eq.
(\ref{state-final-approx}) by the refraction that occurs at the
crystal surface. In this subsection, we neglect the effect of this
refraction for simplicity. This effect is considered in Section
\ref{sec:image-magnification}, where we determine the image
magnification.
\bibitem {Note-bs-output} When constructive interference is observed in
one of the outputs of the beam splitter, destructive interference
occurs in the other output. Because of this changing values of
$\phi_P$ from $\phi_{PC}$ to $\phi_{PD}$ can be avoided during the
experiment by collecting both outputs of the beam splitter. However,
it must be noted that coincidence counts are never detected in this
experiment. Detection of coincidence counts are required for
investigating higher order interference phenomena (see, for example,
\cite{WZM-ind-coh-PRA,BKKC-pdc-fourth-interf}).
\bibitem {Note-img-ninv-lens} It is to be noted that by changing the
lens system in the experimental setup, it is possible to change the
relation between $\widetilde{\k}_I$ and $\k_I$ but not the phase
matching condition. In principle, it is thus possible to obtain an
inverted image by using a different lens system.
\bibitem {Note-ind-em} The higher order terms present in
$\ket{\psi_1}$ and $\ket{\psi_2}$ contain higher (4 or more) photon
states. Because of this the effect of induced emission can be
observed in certain phenomena (see, for example,
\cite{LSGHOG-ind-emiss}) where the lower photon-number states do not
contribute (i.e., one can no longer neglect the higher order terms).
\end{thebibliography}
\end{document}